\documentclass[journal,twocolumn]{IEEEtran}
\usepackage{cite}
\usepackage{amsmath,amsfonts,amsxtra,amssymb,latexsym,amscd,amsthm,mathrsfs,bm}
\usepackage{tabularx}
\usepackage{graphicx}
\usepackage{hyperref}
\usepackage{xcolor}
\usepackage{algorithm}
\usepackage{algorithmic}
\usepackage[colorinlistoftodos,textsize=tiny]{todonotes}
\usepackage{balance}
\usepackage[belowskip=-10pt,aboveskip=0pt,small]{caption}

\newcommand{\snr}{\overline{\mathrm{SNR}}}

\usepackage{eqparbox}

\usepackage{etoolbox}  
\makeatletter
\patchcmd{\algorithmic}{\addtolength{\ALC@tlm}{\leftmargin} }{\addtolength{\ALC@tlm}{\leftmargin}}{}{}
\makeatother

\usepackage{epstopdf}					
\title{\LARGE Performance Improvement of LoRa Modulation with Signal Combining and Semi-Coherent Detection
\thanks{The authors are with the Department of Electrical and Computer Engineering, University of Saskatchewan, Saskatoon, Canada S7N5A9. Emails: \{khai.nguyen, ha.nguyen, e.bedeer\}@usask.ca.}
}

\author{\IEEEauthorblockN{The Khai Nguyen, Ha H. Nguyen, and Ebrahim Bedeer}}

\begin{document}
\maketitle

\begin{abstract}
In this paper, we investigate performance improvements of low-power long-range (LoRa) modulation when a gateway is equipped with multiple antennas. We derive the optimal decision rules for both coherent and non-coherent detections when combining signals received from multiple antennas. To provide insights on how signal combining can benefit LoRa systems, we present expressions of the symbol/bit error probabilities of both the coherent and non-coherent detections in AWGN and Rayleigh fading channels, respectively. Moreover, we also propose an iterative semi-coherent detection that does not require any overhead to estimate the channel-state-information (CSI) while its performance can approach that of the ideal coherent detection. Simulation and analytical results show very large power gains, or coverage extension, provided by the use of multiple antennas for all the detection schemes considered.
\end{abstract}
\vspace*{-0.15cm}

\vspace*{-0.25cm}
\section{Introduction}
Internet of Things (IoT) networks aim to connect a massive number of end devices (EDs) that are typically battery powered and expected to last for several years. Moreover, EDs can be deployed in geographical areas extending to several tens of kilometers and served by a few gateways (GWs) \cite{LongRangeCommunications}. Low-power wide-area networks (LPWANs) are emerging solutions to balance the trade-off among coverage, power requirements and data rates of IoT networks. Among various solutions, low-power long-range (LoRa) technology is currently one of the most widely deployed LPWAN technologies around the world \cite{LongRangeCommunications}. This technology is based on a proprietary chirp spread spectrum (CSS) modulation, known as LoRa modulation, in the PHY layer and LoRaWAN protocol in the MAC layer \cite{LongRangeCommunications}.

With CSS data symbols are modulated into chirp signals whose frequency sweeps
the entire bandwidth once. There are three main parameters in CSS, namely, coding rate, bandwidth (125, 250 or 500 kHz), and spreading factor (from 7 to 12), that can be selected to balance transmission rate, reception sensitivity, and coverage range. In a typical LoRa network, each ED communicates with several \emph{single-antenna} GWs in the uplink transmission. Then, the GWs forward the received signals of each ED, the received signal strength indicator (RSSI) levels, and optional time stamps to the LoRa network server (LNS) through IP backbone. The LNS keeps the received message with the highest RSSI and drops the rest. For downlink transmission to a specific ED, the LNS picks a GW having the highest RSSI from that ED.

It is pointed out that most of the research works on LoRa modulation consider that GWs are equipped with a single antenna. For example, the authors in \cite{LoRaBERApprox} analyze the bit error rate (BER) performance of LoRa modulation and provided tight closed-form approximations in both additive white Gaussian noise (AWGN) and Rayleigh fading channels. In \cite{Hanif-IoTJ20}, the authors introduce a slope-shift-keying LoRa scheme that can increase data rates of the conventional LoRa system by adding a down chirp and its cyclic shifts. They also develop low-complexity optimum coherent and non-coherent detectors, as well as, tight approximations for BER and symbol error rate (SER) when the non-coherent detector is employed in the presence of Rayleigh fading channels. Index modulation is exploited in \cite{Hanif-IoTJ21} to further improve data rates of the conventional CSS/LoRa system and the authors also derive optimal coherent and non-coherent detection rules. Further, they propose a low-complexity non-coherent detection scheme whose performance approaches the optimal performance.

With respect to using multiple antennas, the authors in \cite{xu2019discrete} consider the downlink transmission of a LoRa network in which both GWs and EDs are equipped with multiple antennas and LoRa symbols are also precoded with space-time block coding.
To reduce hardware cost and power consumption at GWs, the number of RF chains is smaller than the number of antennas at each GW, and the authors develop an antenna selection algorithm using discrete particle swarm optimization. Furthermore, a correlated channel environment between the GW and ED, as well as perfect CSI are assumed in \cite{xu2019discrete}.

The first contribution of this paper is to develop signal combining from multiple-antennas at the GW and provide analysis and insights on how such signal combining can benefit LoRa systems. To this end, we evaluate the BER performance of a LoRa system in which each GW is equipped with multiple-antenna and in both cases of coherent detection in an AWGN channel and non-coherent detection in a Rayleigh fading channel. As in the case of $M$-ary FSK with multiple antenna combining \cite{proakis2008digital}, the obtained results show that proper combining of multiple signals received at the GW can significantly improve the LoRa modulation performance, which results in transmit power saving and/or extended coverage for the EDs. The second contribution is a novel iterative semi-coherent detection method that can blindly estimate the CSI, i.e., without using any training overhead, and outperform the non-coherent detection, especially when the number of antennas at the GW increases.

The remainder of this paper is organized as follows. Section \ref{sec:model} introduces the system model. Sections \ref{sec:coh} and \ref{sec:noncoh} present BER expressions of LoRa modulation with multiple receive antennas for AWGN and Rayleigh fading channels, respectively. Section \ref{sec:semicoh} describes the proposed iterative semi-coherent detection method and discusses its computational complexity. Theoretical and simulation results are reported in Section \ref{sec:sim}. Section \ref{sec:con} concludes the paper.

\section{System Model}\label{sec:model}

Consider the uplink transmission of a LoRa network where each ED communicates with a GW equipped with $L$ antennas\footnote{The techniques and results in this paper are also applicable when combining signals received from different single-antenna GWs.}. One LoRa symbol has a bandwidth $B$ and duration $T_{\rm sym}$. Let $T_s = 1/B$ be the sampling period. Then each LoRa symbol can be represented by $M = T_{\rm sym}/T_s = 2^{\textup{SF}}$ samples, where $\textup{SF} \in \{7, 8, \ldots, 12\}$ is the spreading factor, which is also the number of information bits encoded into one LoRa symbol. 
The baseband discrete-time basic symbol (which is an up chirp) of length $M$ samples is given as \cite{Hanif-IoTJ20}:
\begin{equation}
x_0[n]=A\exp\left(j2\pi\left(\frac{ n^2}{2M}-\frac{n}{2}\right)\right), n=0,1\dots, M-1.
\end{equation}
Then, the set of $M$ orthogonal chirps can be simply constructed from $x_0[n]$ as $x_m[n]=x_0[n+m]$, $m=0,1\dots, M$. With the sampling period of $T_s=T_{\rm sym}/M$, the equivalent analog (continuous-time) chirps have an equal energy of $E_s=\int_{0}^{T_{\rm sym}}|x_0(t)|^2 {\rm d}t = \left(\sum_{0}^{M-1}|x_0[n]|^2\right) T_s=A^2 T_{\rm sym}$. It also follows that the signal power is $P_{\rm signal}=\frac{E_s}{T_{\rm sym}}=A^2$.

We consider a frequency-flat and slow Rayleigh fading channel between each ED and each receive antenna. With such a channel model, the received baseband signal corresponding to the $\ell$th antenna of the GW is
\begin{equation}\label{eq3}
{y}_\ell[n]=h_\ell \exp(j\phi) x_{m}[n]+{w}_\ell[n]
\end{equation}
where  ${w}_\ell[n] \sim \mathcal{CN}(0,\sigma^2)$ is a zero-mean AWGN sample at the $\ell$th antenna, $\phi$ is an unknown phase rotation caused by hardware impairments in processing and down converting the received RF signal to baseband, $h_\ell=\alpha_{\ell}\exp\left(j\theta_\ell\right)$ represents the complex channel coefficient ($\alpha_{\ell}$ and $\theta_\ell$ are, respectively, the channel's attenuation and phase shift). For the Rayleigh fading channel, $\alpha_{\ell}$ follows a Rayleigh distribution with average power gain of 1, i.e., $\mathbb{E}\left\{|\alpha_{\ell}|^2\right\}=1$, whereas $\theta_\ell$ is uniformly distributed over $[0,2\pi]$. Obviously, the phase term $\phi$ can be lumped into $\theta_\ell$, or equivalently can be set to zero, and it is done so in the rest of the paper. Note also that the case of an AWGN channel corresponds to setting $\alpha_{\ell}=1$ and $\theta_\ell=0$, $\forall \ell$.

Using the standard notation of $N_0$ for the one-sided power spectral density of the AWGN, the noise power is $P_{\rm noise}= N_0 B$, which is also the variance $\sigma^2$ of the AWGN sample $w_{\ell}[n]$ in \eqref{eq3}. Then, the average signal-to-noise ratio (SNR) at each receive antenna, denoted as $\snr$, is given as $\snr=\frac{P_{\rm signal}}{P_{\rm noise}}=\frac{{E_s}/{T_{\rm sym}}}{N_0 B}=\frac{E_s/M}{N_0}$. This $\snr$ is the same as $A^2/\sigma^2$ calculated based on the discrete-time baseband model in \eqref{eq3}.

\section{Coherent Detection over AWGN Channels} \label{sec:coh}

In coherent detection, perfect knowledge of the CSI is assumed and used to perform the maximal ratio combining (MRC) of $L$ received signals. Specifically, the received signal at the $\ell$th antenna is first correlated with the complex conjugate of the respective channel coefficient as follows \cite{ghanaatian2019lora}:
\begin{equation}\label{eq4}
v_{\ell}^{(\rm coh)}[n] = h_{\ell}^*y_{\ell}[n]= \alpha_{\ell}^2x_m[n]+\hat{w}_{\ell}[n], \quad l = 1, \ldots, L,
\end{equation}
where $\hat{w}_{\ell}[n]\sim\mathcal{CN}\left(0,\alpha_{\ell}^2\sigma^2\right)$. The received signal after the MRC is given as
\begin{equation}\label{eq5}
r[n]=\sum_{\ell=1}^{L}v_{\ell}^{(\rm coh)}[n]=\left(\sum_{\ell=1}^{L}\alpha_{\ell}^2\right)x_m[n]+\hat{\hat{w}}[n],
\end{equation}
where $\hat{\hat{w}}[n] \sim \mathcal{CN}\left(0, \sigma^2 \sum_{l = 1}^{L}\alpha_{\ell}^2 \right)$. To demodulate the combined signal $r[n]$, we perform normalizing and dechirping
as follows
\begin{IEEEeqnarray}{rcl}\label{eq7}
&&z[n] = \frac{r[n]}{\sqrt{\sum_{\ell=1}^{L}\alpha_{\ell}^2}}\frac{x_0^{*}[n]}{A} \nonumber\\
&&=\beta A\underbrace{\exp\left(j2\pi\left(\frac{ m^2}{2M}-\frac{m}{2}\right)\right)}_{\text{constant phase } \Psi_m}\underbrace{\exp\left(\frac{j2\pi mn}{M}\right)}_{\text{linear phase}}+\bar{w}[n], \IEEEeqnarraynumspace
\end{IEEEeqnarray}
where $\bar{w}[n]\sim\mathcal{CN}(0,\sigma^2)$ and $\beta=\sqrt{\sum_{\ell=1}^{L}\alpha_{\ell}^2}$ is the effective gain due to MRC. Then, we perform the $M$-point FFT on the dechirped signal $z[n]$ as
\begin{IEEEeqnarray}{rcl}\label{eq6}
&&Z^{(\rm coh)}[k]= \frac{1}{\sqrt{M}}\sum_{n=0}^{M-1}z[n]\exp\left(\frac{-j2\pi nk}{M}\right) \nonumber\\
&&=\frac{1}{\sqrt{M}}\sum_{n=0}^{M-1}\left(\beta A{\exp\left(j\Psi_m\right)}{\exp\left(\frac{j2\pi mn}{M}\right)}+\bar{w}[n]\right) \nonumber\\
&&\quad{\times\exp\left(\frac{-j2\pi nk}{M}\right)}\nonumber\\
&&=\begin{cases}
\beta A\sqrt{M}\exp\left(j\Psi_m\right)+W[m],\quad k=m\\
W[k],\quad k = 1, \ldots, M, \: k\neq m,
\end{cases}
\end{IEEEeqnarray}
where $W[\cdot] \sim \mathcal{CN}(0,\sigma^2)$ is the noise sample after the FFT. One can see from \eqref{eq7} that the phase $\Psi_m$ is deterministic, hence, the equivalent decision variable can be re-written as
\begin{equation}\label{eq:dv0}
\begin{split}
Z_{\rm R}^{(\rm coh)}[k]&=\mathfrak{R}\left\{Z^{(\rm coh)}[k]\exp\left(-j\Psi_m\right)\right\}\\
&=\begin{cases}
\beta A\sqrt{M}+W_{\rm R}[m],\quad k=m\\
W_{\rm R}[k],\quad k = 1, \ldots, M, \: k\neq m,
\end{cases}
\end{split}
\end{equation}
where $\mathfrak{R}\left\{\cdot\right\}$ returns the real value and $W_{\rm R}[\cdot] \sim \mathcal{CN}(0,\sigma^2/2)$. For ease in performance analysis, we scale the decision variable in \eqref{eq:dv0} by $\sqrt{B}$ to obtain $\hat{Z}_{\rm R}^{(\rm coh)}[k]=Z_{\rm R}^{(\rm coh)}[k]/\sqrt{B}$. Using the relationships $E_s=A^2 T_{\rm sym}=A^2 M/B$ and $\sigma^2 =N_0 B$, the scaled decision variable is expressed as
\begin{equation}\label{eq:dv}
\hat{Z}_{\rm R}^{(\rm coh)}[k]=\begin{cases}
\beta \sqrt{E_s}+\hat{W}_{\rm R}[m],\quad k=m\\
\hat{W}_{\rm R}[k],\quad k = 1, \ldots, M, \: k\neq m,
\end{cases}
\end{equation}
where $\hat{W}_{\rm R}[\cdot] \sim \mathcal{CN}(0,N_0/2)$. Finally, the decision rule is given as
\begin{equation}\label{eq10}
\hat{m}^{(\rm coh)}=\underset{k=0,1,\dots, M-1}{\text{arg max}}\hat{Z}_{\rm R}^{(\rm coh)}[k].
\end{equation}

Coherent detection is not very practical for fading channels since it needs precise channel estimation (which typically comes with extra training overhead and complexity). As such, we complete this section by presenting the error probability of LoRa modulation with MRC in AWGN channels, i.e., when $\alpha_{\ell}=1$ and $\theta_{\ell}=0$, and $\beta=\sqrt{L}$. One can observe that the decision variable in \eqref{eq:dv} is the same as that for orthogonal $M$-ary FSK. Therefore, the error probability of LoRa modulation with the maximum-ratio combining of $L$ signals is the same as in the single antenna case for orthogonal modulations, except with a power gain of $L$ due to the MRC. Accordingly, the symbol error probability for the LoRa modulation can be obtained by replacing $E_s$ with $LE_s$ in the symbol error probability of $M$-ary FSK using a single antenna \cite[equation 8.67]{nguyen2009first}. This yields
\begin{IEEEeqnarray}{rcl}\label{eq12}
&&P_s^{(\rm coh)}=1-\frac{1}{\sqrt{2\pi}}\int_{-\infty}^{\infty}\left(\frac{1}{\sqrt{2\pi}}\int_{-\infty}^{y}\exp\left(\frac{-x^2}{2}\right)dx\right)^{M-1} \nonumber\\
& &\times\exp\left(-\frac{1}{2}\left(y-\sqrt{\frac{2 L E_s}{N_0}}\right)^2\right)dy.
\end{IEEEeqnarray}
Finally, the bit error probability of LoRa modulation in AWGN channels with MRC can be found as $P_b^{(\rm coh)}=\frac{M}{2(M-1)}P_s^{(\rm coh)}$.

\section{Non-Coherent Detection over Fading Channels}\label{sec:noncoh}

Non-coherent detection of LoRa modulation is much more relevant for IoT networks in practice as it does not require precise knowledge of the CSI. The first step in non-coherent detection is to perform normalizing and dechirping of the received signal at the $\ell$th antenna in \eqref{eq3} as follows
\begin{IEEEeqnarray}{rcl}\label{eq:non1}
&&{v}_{\ell}^{(\rm non-coh)}[n] = y_{\ell}[n] \frac{x_0^*[n]}{A}, \nonumber\\
&&=\alpha_{\ell}\exp\left(j\theta_{\ell}\right)A {\exp\left(j\Psi_m\right)}{\exp\left(\frac{j2\pi mn}{M}\right)}+\bar{w}_{\ell}[n],\IEEEeqnarraynumspace
\end{IEEEeqnarray}
where $\bar{w}_{\ell}[n] \sim \mathcal{CN}(0,\sigma^2)$. Then by performing the $M$-point FFT on the dechirped signal $v_{\ell}^{(\rm non-coh)}[n]$ and dividing the result by $\sqrt{B}$, we obtain
\begin{IEEEeqnarray}{rcl}\label{eq19}
V_{\ell}[k]&{}={}&\begin{cases}
\sqrt{E_s}\alpha_{\ell}\exp\left(j\theta_{\ell}\right)\exp\left(j\Psi_m\right)+W_{\ell}[m],\: k=m \nonumber\\
W_{\ell}[k],\quad k = 1, \ldots, M, \:k \neq m
\end{cases}\\
&{}={}&\begin{cases}
\sqrt{E_s} a_{\ell}[m] +W_{\ell}[m],\quad k=m \\
W_{\ell}[k],\quad k = 1, \ldots, M, \:k \neq m,
\end{cases}
\end{IEEEeqnarray}
where $a_{\ell}[k] = \alpha_{\ell}\exp\left\{j\theta_{\ell}\right\}\exp\left\{j\Psi_k\right\}=h_{\ell}\exp\left\{j\Psi_k\right\} \sim\mathcal{CN}(0,1)$ and $W_{\ell}[k]\sim\mathcal{CN}(0,N_0)$.

Since CSI is not available in non-coherent detection, we apply the square-law combining to all the received  signals from the $L$ antennas at bin $k$ as
\begin{eqnarray}\label{eq:dev-non-coh}
&&Z^{(\rm non-coh)}[k]=\sum_{\ell=1}^{L}\left\vert V_{\ell}[k]\right\vert^2\nonumber\\
&&=
\begin{cases}
\sum_{\ell=1}^{L}\left\vert \sqrt{E_s}a_{\ell}[m]+W_{\ell}[m]\right\vert^2,\: k=m\\
\sum_{\ell=1}^{L}\left\vert W_{\ell}[k]\right\vert^2,\quad k = 1, \ldots, M, \:k \neq m.
\end{cases}
\end{eqnarray}
Then, the final decision rule can be expressed as
\begin{equation}\label{eq:non-coh}
\hat{m}^{(\rm non-coh)}=\underset{k=0,1,\dots, M-1}{\text{arg max}}Z^{(\rm non-coh)}[k].
\end{equation}

To find the error probability for the non-coherent detection of LoRa modulation, we follow a similar analysis as in \cite{proakis2008digital}. Without loss of generality, if we assume $x_0[n]$ was transmitted, then the square-law combining output at the 0th bin is $\Lambda_0=Z^{(\rm non-coh)}[0]=\sum_{\ell=1}^{L}\left\vert \sqrt{E_s}a_{\ell}[0]+W_{\ell}[0]\right\vert^2$, whereas the output at bin $k$, $\forall k$, $k \ne 0$ is $\Lambda_k=Z^{(\rm non-coh)}[k]=\sum_{\ell=1}^{L}\left\vert W_{\ell}[k]\right\vert^2$. To calculate the probability of error, we need first to find the probability that $\Lambda_0 > \Lambda_k$, $\forall k$, $k \ne 0$. One can show that $\Lambda_k$, $\forall k$, are mutually statistically independent random variables. Hence,
\begin{equation}\label{eq:eq123}
\begin{split}
&P(\Lambda_1<\Lambda_0,\Lambda_2<\Lambda_0,\dots, \Lambda_{M-1}<\Lambda_0)\\
&=\left(P(\Lambda_1<\Lambda_0)\right)^{M-1}=\left(\int_{0}^{\lambda_0}f(\lambda_1)d\lambda_1\right)^{M-1}\\
&=\left(1-\exp\left(-\frac{\lambda_0}{\sigma_1^2}\right)\sum_{q=0}^{L-1}\frac{1}{q!}\left(\frac{\lambda_0}{\sigma_1^2}\right)^q\right)^{M-1},
\end{split}
\end{equation}
where $f(\lambda_1)$ is the probability density function of $\Lambda_1$ given~as
\begin{equation}
f(\lambda_1)=\frac{1}{(\sigma_1^2)^{L}(L-1)!}\lambda_1^{L-1}\exp\left(-\frac{\lambda_1}{\sigma_1^2}\right),
\end{equation}
and $\sigma_1^2=\mathbb{E}\left\{\left\vert W_{\ell}[k]\right\vert^2\right\}=N_0$. The probability of a correct symbol decision can be calculated by averaging \eqref{eq:eq123} over the probability density function of $\Lambda_0$, which is
\begin{equation}
f(\lambda_0)=\frac{1}{(\sigma_0^2)^{L}(L-1)!}\lambda_0^{L-1}\exp\left(-\frac{\lambda_0}{\sigma_0^2}\right),
\end{equation}
where $\sigma_0^2=\mathbb{E}\left\{\left\vert \sqrt{E_s}a_{\ell}[m]+W_{\ell}[0]\right\vert^2\right\}=E_s+N_0$. Then, the probability of symbol error equals 1 minus the probability of a correct symbol decision and is given as \cite[equation 14.4--47]{proakis2008digital}
\begin{eqnarray}\label{eq29}
&&P_s^{(\rm non-coh)}=1-\int_{0}^{\infty}\frac{1}{(\sigma_0^2)^{L}(L-1)!}\lambda_0^{L-1}\exp\left(-\frac{\lambda_0}{\sigma_0^2}\right)\nonumber\\
&&\times\left(1-\exp\left(-\frac{\lambda_0}{\sigma_1^2}\right)\sum_{q=0}^{L-1}\frac{1}{q!}\left(\frac{\lambda_0}{\sigma_1^2}\right)^q\right)^{M-1}d\lambda_0\nonumber\\
&&=1-\int_{0}^{\infty}\frac{1}{(1+\bar{\gamma}_c)^{L}(L-1)!}\lambda_0^{L-1}\exp\left(-\frac{\lambda_0}{1+\bar{\gamma}_c}\right)\nonumber\\
&&\times\left(1-\exp\left(-{\lambda_0}\right)\sum_{q=0}^{L-1}\frac{\lambda_0^q}{q!}\right)^{M-1}d\lambda_0,
\end{eqnarray}
where $\bar{\gamma}_c=E_s/N_0=M \snr$. In Section \ref{sec:sim}, we evaluate \eqref{eq29} directly using numerical integration. Finally, the bit error probability of non-coherent detection of LoRa modulation with signal combining is given as $P_b^{(\rm non-coh)}=\frac{M}{2(M-1)}P_s^{(\rm non-coh)}$.

\section{Proposed Semi-Coherent Detection}\label{sec:semicoh}

In this section, we propose an iterative \emph{semi-coherent} detector for LoRa modulation when combining $L$ received signals without needing extra overhead for CSI estimation.  Section \ref{sec:sim} shows that the error performance of the proposed detector can closely approach that of the ideal coherent detector.

Assume that the fading channels stay constant for a coherence time of $\tau_c$ LoRa symbols. In the first stage, initial detection of $\tau_c$ LoRa symbols is carried out using the non-coherent detector in \eqref{eq:non-coh}. Let $\hat{\boldsymbol{m}}=[\hat{m}_1,\ldots,\hat{m}_{\tau_c}]$ be a vector containing these initial detected LoRa symbols. Suppose that a specific LoRa symbol, say the $i$th symbol is detected correctly by the non-coherent detector. Then it follows from \eqref{eq19} that the least-square estimate of the channel coefficient $h_{\ell}$ can be obtained as $\hat{h}_{\ell,i}=V_{\ell,i}[\hat{m}_i]\exp\left(-j\Psi_{\hat{m}_i}\right)/\sqrt{E_s}$, where $V_{\ell,i}[\hat{m}_i]$ is the FFT value at the $\hat{m}_i$th bin over the $i$th symbol duration, and the symbol-dependent phase rotation $\Psi_{\hat{m}_i}$ can be simply computed (see \eqref{eq7}) as  $\Psi_{\hat{m}_i} = 2\pi \left(\frac{\hat{m}_i^2}{2M} - \frac{\hat{m}_i}{2}\right)$.

Obviously $\hat{h}_{\ell,i}$ is a good estimate of the true channel coefficient $h_{\ell}$ if $\hat{m}_i=m_i$, whereas it contains only noise when $\hat{m}_i\neq m_i$. Fortunately, since the channel $h_{\ell}$ stays constant over $\tau_c$ LoRa symbols, one can obtain a good channel estimate by averaging $\hat{h}_{\ell,i}$ over all $\tau_c$ symbols to obtain:
\begin{equation}\label{eq23}
\hat{h}_{\ell}^{(\mathrm{ave})}=\frac{1}{\tau_c}\sum_{i=1}^{\tau_c}\hat{h}_{\ell,i}
=\frac{1}{\sqrt{E_s}\tau_c}\sum_{i=1}^{\tau_c}V_{\ell,i}[\hat{m}_i]\exp\left(-j\Psi_{\hat{m}_i}\right).
\end{equation}
Note that, because the noise components $W_{\ell,i}[k]$ of $V_{\ell,i}[k]$ are independent from symbol to symbol (see \eqref{eq19}), when the coherent time goes to infinity, we can obtain the exact channel value, i.e., $\hat{h}_{\ell}^{(\mathrm{ave})}\xrightarrow[\tau_c\rightarrow\infty]{\mathrm{a.s}} h_{\ell}$.

In the second stage, we make use of the channel estimate $\hat{h}_{\ell}^{(\mathrm{ave})}$ to perform new detection of all $\tau_c$ LoRa symbols. Specifically, for making a new detection of the $i$th symbol, we use $\hat{h}_{\ell}^{(\mathrm{ave})}$ to co-phase $V_{\ell,i}[k]$ before coherently combining (adding) the signals received from all $L$ antennas. The decision variable for the proposed semi-coherent detection is
\begin{equation}\label{eq:dev-semi-coh}
Z_i^{(\rm semi-coh)}[k]=\sum_{\ell=1}^{L}\left(\hat{h}_{\ell}^{(\mathrm{ave})}\exp\left(j\Psi_k\right)\right)^{*} V_{\ell,i}[k],
\end{equation}
and the final decision rule can be expressed as
\begin{equation}\label{eq:max-semi-coh}
\hat{m}_i^{(\rm semi-coh)}=\underset{k=0,\dots, M-1}{\text{arg max}}\mathfrak{R}\left\{Z_i^{(\rm semi-coh)}[k]\right\},\; i=1,\ldots,\tau_c.
\end{equation}
It is pointed out that, since the decision rule in \eqref{eq:max-semi-coh} is obtained by finding the maximum of $\mathfrak{R}\{Z_i^{(\rm semi-coh)}[k]\}$, it is invariant to the scaling of the channel estimate $\hat{h}_{\ell}^{(\mathrm{ave})}$, and hence the scaling factor in front of the summation in \eqref{eq23} can be ignored.

After this new detection operation, a new average channel estimate is obtained, while the newly detected LoRa symbols are saved and compared to the previously-detected results. The iteration process between detection and estimation continues until convergence, i.e., when the detected symbols in two consecutive iterations are the same. When operating at low SNRs, the iterative semi-coherent detector does not likely converge because the initial non-coherent detection is not reliable and the channel estimation is largely based on the wrong frequency bins. In such a case, we use a predetermined number of iterations, i.e., $ N_{\mathrm{max}}$, as a stopping criterion. The proposed iterative semi-coherent detection algorithm is summarized in Algorithm~\ref{al:al2}.

\begin{algorithm}
	\caption{Iterative semi-coherent detection}
	{\small
		\begin{algorithmic}[1]\label{al:al2}
			\REQUIRE Spreading factor (SF), $\tau_c$, $N_{\mathrm{max}}$.
			\STATE Initially detect $\tau_c$ symbols $\hat{\boldsymbol{m}}=[\hat{m}_1,\ldots,\hat{m}_{\tau_c}]$ with non-coherent detection.
			\STATE Estimate $\hat{h}_{\ell,i}$ corresponding to $\tau_c$ detected LoRa symbols ($i=1,2,\dots,\tau_c$).
			\STATE Obtain the average channel estimate $\hat{h}_{\ell}^{(\mathrm{ave})}$ as in \eqref{eq23}.
			\STATE Save current decision as $\hat{\boldsymbol{m}}_{\mathrm{prev}}=\hat{\boldsymbol{m}}$.
			\STATE $\mathrm{flag}=1$; $\mathrm{count=0;}$
			\WHILE{$\mathrm{flag}=1$ $\&$ $\mathrm{count}\leq N_{\mathrm{max}}$}
			\STATE $\mathrm{count}=\mathrm{count}+1;$
			\STATE For the $i$th symbol, combine $V_{\ell,i}[k]$ with $\hat{h}_{\ell}^{(\mathrm{ave})}\exp\left(j\Psi_{k}\right)$ as in \eqref{eq:dev-semi-coh} and perform a new detection of the $i$th symbol $\hat{m}_i$ as in \eqref{eq:max-semi-coh}, $\forall i=1,2,\dots,\tau_c$.
			\STATE Update $\hat{h}_{\ell}^{(\mathrm{ave})}$ using newly detected symbols $\hat{\boldsymbol{m}}$.
			\IF {$\hat{\boldsymbol{m}}_{\mathrm{prev}}\neq\hat{\boldsymbol{m}}$}
			\STATE $\mathrm{flag}=1$
			\ELSE
			\STATE $\mathrm{flag}=0$
			\ENDIF
			\STATE $\hat{\boldsymbol{m}}_{\mathrm{prev}}=\hat{\boldsymbol{m}}$
			\ENDWHILE
			\RETURN $\hat{\boldsymbol{m}}$.
		\end{algorithmic}
	}
\end{algorithm}

\textit{Complexity Analysis:} As discussed earlier, the first step of the semi-coherent detection is to perform non-coherent detection. Hence, to evaluate the computational complexity of the proposed semi-coherent detection, we need to evaluate the complexity of the non-coherent detection first.

For the non-coherent detection, evaluating \eqref{eq:non1} for all $L$ antennas requires complexity in the order of $\mathcal{O}(L)$ operations. Performing the $M$-point FFT in \eqref{eq19} requires complexity of $\mathcal{O}(M \log_2 M)$ operations per antenna, or  $\mathcal{O}(L M \log_2 M)$ operations for all $L$ antennas. The square-law combining in \eqref{eq:dev-non-coh} requires complexity of $\mathcal{O}(L)$ operations, and evaluating the decision rule in \eqref{eq:non-coh} requires complexity of $\mathcal{O}(M)$ operations. Hence, it is simple to see that the overall computational complexity of the non-coherent detection to detect one LoRa symbol is in the order of $\mathcal{O}(L M \log_2 M)$ operations.

For the semi-coherent detection, estimating $\hat{h}_{\ell,i}$ in Step 2 of Algorithm~\ref{al:al2} for all $\tau_c$ symbols requires complexity of $\mathcal{O}(\tau_c)$ operations. Also, calculating $\hat{h}_{\ell}^{(\mathrm{ave})}$ in Step 3 according to \eqref{eq23} requires complexity of $\mathcal{O}(\tau_c)$ operations. One can show that Step 8 requires complexity of $\mathcal{O}(L \tau_c)$ and $\mathcal{O}(M \tau_c)$ operations in order to evaluate \eqref{eq:dev-semi-coh} and \eqref{eq:max-semi-coh} for $\tau_c$ symbols, respectively. Since the \textbf{while} loop in Step 6 repeats at most $N_{\mathrm{max}}$ times, the complexity of Steps 6 to 16 is $\mathcal{O}(\tau_c N_{\mathrm{max}}  (L + M))$. Given that $\tau_c < M$, the overall computational complexity of the semi-coherent detection to detect all $\tau_c$ symbols is  $\mathcal{O}(\tau_c L M \log_2(M) +\tau_c N_{\mathrm{max}} (L + M))$ operations.

\section{Results and Discussion}\label{sec:sim}
In this section, we provide numerical results for the BER of both coherent detection in an AWGN channel and non-coherent detection in a Rayleigh fading channel to validate our theoretical analysis. We also show the BER performance of the proposed iterative semi-coherent detection. A minimum of $10^5$ Monte-Carlo trails were used to generate the BER curves for the detectors under study and $ N_{\mathrm{max}}$ is set to 50. For the Rayleigh channel, it is assumed to be frequency flat and stays constant over the duration of $\tau_c = 10$ LoRa symbols.

\begin{figure}[t!]
	\centering
	\includegraphics[width=0.475\textwidth]{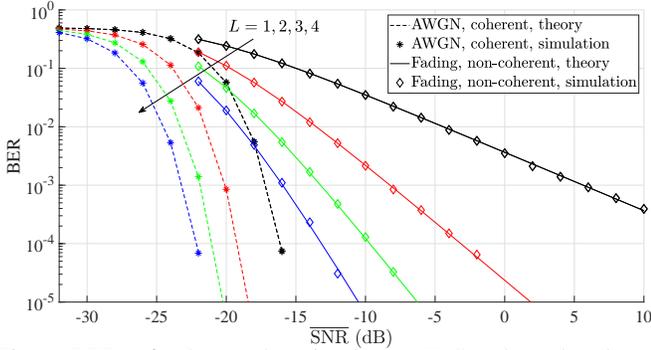}
	\caption{BERs of coherent detection in an AWGN channel and non-coherent detection in a Rayleigh fading channel: ${\rm SF} = 10$ and $L = 1, 2, 3, 4$.}
	\label{fig:Coh_Noncoh_SF_10}
\end{figure}

Fig. \ref{fig:Coh_Noncoh_SF_10} plots the simulated and theoretical BER curves of both coherent detection in AWGN channels and non-coherent detection in Rayleigh fading channels versus the average received signal-to-noise ratio per antenna, i.e., $\snr$,
and for ${\rm SF} = 10$. As one can observe, the simulated BER results of both detection techniques match perfectly with the theoretical results. As expected, increasing $L$ helps to improve the BER performance. In particular, for a BER of $10^{-4}$, increasing $L$ from 1 to 4 results in approximately 8 and 29 dB savings in the SNR (i.e., the transmit power) for the coherent and non-coherent detection cases, respectively. Considering a typical LoRa free-space path loss model of $91.22+10 n \log(d/d_0)$ (dB) and with the reference distance $d_0=1$ km and path-loss exponent $n=2$ \cite{petajajarvi2015coverage}, the SNR savings of 8 and 29 dB translate, respectively, to 2.5 times and 27 times larger in coverage extension for a fixed transmit power of the ED.

\begin{figure}[t!]
	\centering
	\includegraphics[width=0.475\textwidth]{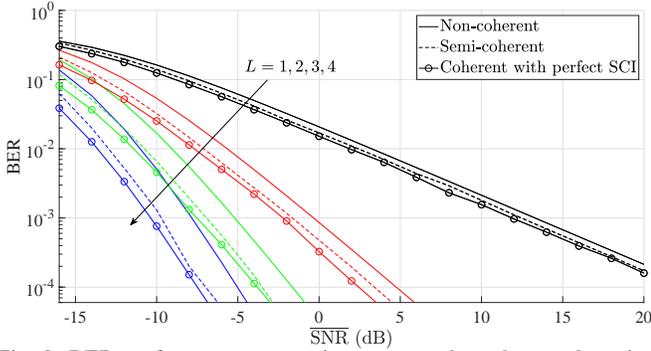}
	\caption{BER performance comparison among the coherent detection, non-coherent detection, and semi-coherent detection in Rayleigh fading channels for ${\rm SF} = 7$, $\tau_c = 10$, and $L = 1, 2, 3, 4$.}
	\label{fig:Semi_vs_L_SF7_tau10}
\end{figure}

Fig. \ref{fig:Semi_vs_L_SF7_tau10} compares the BER performance of the non-coherent (by theory), semi-coherent (by simulation), and coherent detection (by simulation) versus $\snr$ over Rayleigh fading channels for ${\rm SF} = 7$ and $\tau_c = 10$.
One can see that the iterative semi-coherent detection (that blindly estimates the CSI) performs within a fraction of dB from the coherent detection (that assumes perfect CSI knowledge). This clearly shows the promise of using the proposed iterative semi-coherent detection in LoRa modulation with signal combining.

\begin{figure}[t!]
	\centering
	\includegraphics[width=0.475\textwidth]{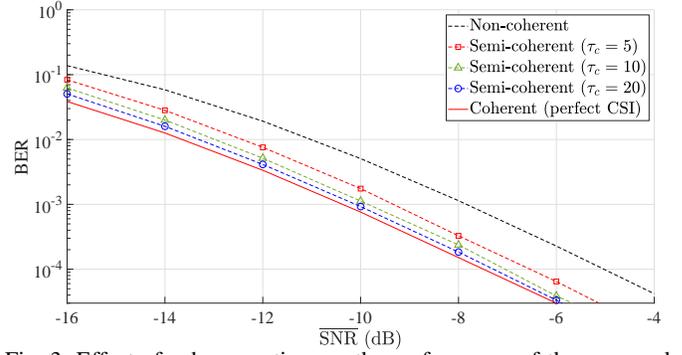}
	\caption{Effect of coherence time on the performance of the proposed semi-coherent detection.}\label{fig:Semi_vs_tau}
\end{figure}

Fig. \ref{fig:Semi_vs_tau} depicts the influence of the coherence time on the BER performance of the proposed semi-coherent detection over a Rayleigh fading channel for ${\rm SF} = 7$, $L=4$, and $\tau_c= 5, 10, 20$ LoRa symbols. As expected, as the channels change slower, i.e., when the coherence time is larger, the blind channel estimate becomes more accurate, and hence the performance of the semi-coherent detection approaches closer to that of the coherent detection. For the worst situation considered, i.e., when $\tau_c= 5$, the semi-coherent detection can still provide an impressive SNR gain of more than 1 dB over the non-coherent detection.

\vspace*{-0.25cm}

\section{Conclusion}\label{sec:con}
We have investigated performance improvements of LoRa modulation when multiple antennas are employed at the GW and by combining signals received over these antennas. The optimal decision rules were established and the corresponding BER expressions were obtained for both coherent and non-coherent detections in AWGN and Rayleigh fading channels, respectively. More importantly, we also proposed an iterative semi-coherent detection whose performance approaches that of the coherent detection without the need to spend extra resources for CSI training or estimation. Simulation results showed that increasing the number of antennas at the GW from 1 to 4 results in approximately 8 and 29 dB savings in transmit power of the ED (which translate to 2.5 times and 27 times larger distance for the same transmit power of the ED), when the system operates over AWGN (coherent) and Rayleigh fading (non-coherent) channels, respectively. The results revealed that the proposed semi-coherent detection performs within a fraction of dB from the coherent detection.

\vspace*{-0.25cm}

\bibliographystyle{ieeetr}

\end{document}